\documentstyle[aps,preprint,epsf]{revtex}
\renewcommand{\ref}[1]{\raisebox{.6ex}{[#1]}}

\newcommand{\be}{\begin{equation}}
\newcommand{\ee}{\end{equation}}

\newcommand{\ba}{\begin{array}}
\newcommand{\ea}{\end{array}}

\begin{document}


\title{ Vortex Dynamics, Resistivity Formula, and
Fluctuation-dissipation Theorems
\footnote{ part of the talk given by XMZ 
 in the 1997 Taiwan conference on superconductivity, Aug. 13-16,
 (Chinese Journal of Physics v.36, 197 (1998)(Taipei) )}}

\author{X.-M. Zhu$^a$ and  P. Ao$^b$   \\
       Department of Experimental$^a$/Theoretical$^b$ Physics \\
       Ume\aa{\ }University, S-901 87, Ume\aa, SWEDEN }

\maketitle

\begin{abstract}
We investigate the problem of forces on moving vortex
in a superfluid or
superconductor. The main purpose is to locate the
source which leads to  the contradictory  results in the literature.
We establish the connection between this problem  to the 
difficult but well studied subject of resistivity formula in transport theory.
The relaxation time
approximation used in the force calculation via 
force-force correlation function
is shown to be invalid. 
The roles of Berry's phase and  fluctuation-dissipation theorems are discussed.
\end{abstract}

For a vortex moving in a superfluid or superconductor, the only
agreement we have reached so far is at zero temperature, 
in absence of any impurity potentials. 
Question arises in more complicated cases.
Using the Berry's phase calculation \cite{at}, as well as the exact
total force-force correlation calculation\cite{thouless},  
It has been shown that the transverse force is the Magnus force,
coming from extended states with no dependence on the core level spacing
and other extrinsic details.
Other microscopic derivations  
using core state transitions\cite{volovik,simanek,kp2,vo,stone}
have shown, however,  
that the transverse force on a moving vortex is greatly reduced
in magnitude when the core level spacing is less than the
energy scale associated with random scattering.
Apparently, it seems  that they
are not two equivalent ways to obtain the same 
quantity but rather at least one of them must be incorrect or incomplete. 
Thus if we could settle the difference on
the transverse force on moving vortex between core or extended states
originated, we should be able to reach an agreement.
However, these two ways of calculating the transverse force on
moving vortex 
are shown to be equivalent to each other\cite{az}. 
It is impossible 
for them to give different results unless there is a
hidden mistake elsewhere.

The algebra in the derivation of vortex dynamics 
is either involved or abstract.  The repeating of 
those calculations is not the most efficient route
to detect and understand the source of disagreements.
Instead, we note that
the vortices satisfy a classical Langevin equation with 
parameters determined microscopically.  
This equation has the same form as a classical electron moving in a
magnetic field.  Therefore general properties 
of Langevin dynamics, for instance the fluctuation-dissipation theorems,
should be obeyed by vortices. In addition,  
our knowledge on electronic transport can be
readily used to help understanding vortex dynamics.
In particular, we will examine the relaxation time approximation 
and show it to be invalid in 
resistivity calculation via force-force correlations. 

Let us consider a classical charged particle in a magnetic
field  obeying a generalized
Langevin equation:
\be
  \ba{cl}
   m \dot{u}_i(t) = & -\int^{t}_{t_0} dt' \; \eta_{ik}(t-t')  u_k(t')
        + { K}_i(t)  \\
        & + B \epsilon_{ik} \; u_k(t) + f_i(t) \; . \\
  \ea
\ee
Here the index $i = x$ or $y$,
the velocity of the particle ${\bf u}(t) = ( u_x(t), u_y(t) ) $, $m$
the mass, ${\bf K}(t)= ( K_x(t), K_y(t) ) $ an external force, 
${\bf f}(t) = ( f_x(t),  f_y(t) )$ a random
force which simulates the effect of the thermal reservoir.
The Einstein convention of the repeated indices as summation has been used.
$B \epsilon_{ik} \; u_k(t)$
represents the transverse force, the Lorentz force
$ {\bf u}(t) \times {\bf B}$ in the Langevin equation with
the magnetic field taken along the $z$-direction. 
The matrix $\eta(t-t') = \{ \eta_{ij} \} $ represents frictions in both 
longitudinal and transverse directions of the particle motion.
Its possible finite off-diagonal elements
will change the effect of the original Lorentz force on the particle.
In addition, we have 
\be
 \begin{array}{ccl}
   < f_i (t)> & =  & 0 \; ,\\
   <  u_i (t_0) f_j (t_0 + t )> & = & 0, \, \, t > 0        \; ,\\
   <u_i(t_0) u_j(t_0) > & =  &  \frac{k_B T}{m} \delta_{ij} \; .   \\
\end{array}
\ee
The second one is due to the causality and the
last one is the equipartition theorem.

The problem of particle response to a perturbation
can  be formulated into two different ways. 
We can calculate the velocity of the particle while the
applied force is given. 
Here the Hamiltonian of the particle is known.
In such a case, 
it is equivalent to obtaining a conductivity formula. 
Otherwise we can consider a given trajectory of particle motion and
calculate what is the applied force needed to maintain it.
It is equivalent to
obtaining a resistivity formula, i.e. calculating electric
field needed to maintain the given current.
The derivation of vortex dynamics  belongs to the
second kind, where we assume a steady motion of vortex and calculate
what is the external force acted on the vortex.
In electron transport, we have a choice to formulate the problem
in either way. In vortex dynamics, we can only formulate in
terms of resistivity because the effective vortex Hamiltonian is unknown 
and is exactly what we want to obtain. 

The conductivity may be  obtained by the Nakano-Kubo's formula, 
a calculation of velocity-velocity correlation function.
It may also be obtained by solving  Boltzmann equation
in presence of an electric field\cite{hc}.
Both of them are standard methods. The relaxation time
approximation is a valid one in  these cases.

The resistivity formula is much more confusing.
Since 1960s, various resistivity formulae have been proposed and
examined repeatedly\cite{kubo,bh,hc,fm}.
In the derivation of vortex dynamics, 
total  force-force correlation
formalism has been explicitly shown by \u{S}im\'{a}nek to be the one 
used in various Green's function calculations of forces on
moving vortex\cite{simanek}. 
In addition, the force-balance theory has  been used to calculate
forces acting on a vortex\cite{kp2}. Recently, Boltzmann equation 
has also been used\cite{stone}. 
In  the literature of electron transport,
all of the above methods have  been used in obtaining a resistivity formula, 
although normally discussed without magnetic field. 
It has been demonstrated that the longitudinal 
resistivity calculated 
directly using  total force-force correlation function is always zero
in the zero frequency limit \cite{bh}.
Thus any formula that allows to 
obtain finite friction, or longitudinal resistivity, directly 
from total force-force correlation function must be incorrect
in this limit. 
The longitudinal resistivity formula using Boltzmann equation gives either
$0/0$ results or a formula trivially equivalent to reciprocal
conductivity formula\cite{hc}. 

We will first briefly review conductivity formula. 
As expected, we will show that in
such a case $<{\bf u}(t)>$
is determined by velocity-velocity correlation functions
in the absence of  ${\bf K}$.  We will  verify that
indeed the relaxation time approximation is 
valid in such a calculation.
Parallelly, we  will derive a resistivity formula by
assuming a given  velocity  $ {\bf u}(t) $ and calculating
the average force $<{\bf K}>$ which should be applied to sustain such
a motion. We find that the problems involved in vortex dynamics
become clear after careful study of this simple model.

Now let us put ${\bf K}= 0$ in Eq.(1) and calculate the velocity-velocity
correlation function. Multiply Eq.(1) by  $ u_j(t_0) $
and take the ensemble average:
\[
 \begin{array}{l}
  m <\dot{u}_i(t_0 + t)  u_j(t_0)>  = \\
  - \int^{t}_{0}dt' \; \eta_{ik} (t-t')< u_k (t_0+t') u_j (t_0)> \\
    + B \epsilon_{ik} < u_k (t_0 + t)  u_j(t_0)>
  + < f_i (t_0 + t)  u_j (t_0)> \, . \\
\end{array}
\]
The term $ < u_i (t_0) f_j(t_0 + t)>$ vanishes according to Eq.(2). 

Introducing a Laplace transform
\[
  \eta[\omega] = \int^{\infty}_{0}  dt \; e^{-i\omega t}\eta(t)  \, ,
\]
defining the velocity-velocity correlation function matrix
\be
   {\cal U}_{ij}(t) =  < u_i(t_0 + t) u_j(t_0) > \; ,  
\ee
and integrating by part, we have 
\be
   {\cal U} [\omega]
   = ( im\omega + \eta[\omega] - i \sigma_y B )^{-1} k_B T   \, .
\ee
We have used the identity $m {\cal U}(0) = k_B T $.
$ \sigma_y = - i \{ \epsilon_{ij} \} $ is the  Pauli matrix.

Next we calculate the total force-force correlation function matrix
\[
  {\cal F}_{ij}(t) = m^2 <\dot{u}_i (t_0 + t) \dot{u}_j (t_0)> \, .
\]
Taking the Laplace transform, using the 
translational invariant in time
\[
  < u_i (t_0+t) \dot{u}_j (t_0)> = -<\dot{u}_i (t_0+t) u_j(t_0)> \; ,
\]
and the total force-velocity correlation function
\[ 
  m <\dot{u}_i (t_0+t) u_j(t_0)>[\omega] = - m {\cal U}_{ij}(0)
     + i m \omega {\cal U}_{ij}[\omega ] \; ,
\]
we have
\be
   \ba{ll}
  {\cal F} [\omega]  = & \left( iB\sigma_y + i m \omega +
  (m\omega)^2\times \right. \\ 
   & \left. \left( im \omega + \eta [\omega]   
      -i \sigma_y B \right) ^{-1}
   \right) k_B T   \, . 
   \ea
\ee
In the limit $\omega\rightarrow 0$, we have
\be
  {\cal F} [0] =  iB\sigma_y \; k_B T  \; ,
\ee
which is independent of $\eta$. 

We then calculate the random force-force correlation matrix
\be
   {\cal R}_{ij}(t) = <f_i (t_0 + t) f_j(t_0)> \; . 
\ee
From Eq.(1) we can express  ${\cal R}(t)$ in terms of total force-force, 
total force-velocity, and velocity-velocity correlation functions.
Taking the Laplace transform and integrating by part, 
we obtain
\be
   {\cal R}[ \omega ] 
   = \eta[\omega] \; m {\cal U}(0) =  \eta[\omega] \; k_B T \, ,
\ee
or $\eta(t) =  {\cal R}(t) /(k_B T ) $.
This is the `second' fluctuation-dissipation theorem described by 
Kubo\cite{kubo}. The generalized friction
$\eta(t)$ is given by the random force-force correlation.
Because the random force is determined by the thermal bath, 
$\eta(t)$ has no off-diagonal part if the random force-force
correlation matrix has not.
An important conclusion we can draw from here is that the
even though there is no time-reversal symmetry in the particle motion,
the frictional force is always longitudinal as long as the heat bath
does not generate an off-diagonal element in the 
random force-force correlation function matrix. 
For example, this is the case for 
a charged particle dynamics 
described by a single relaxation time in Boltzmann equation.

Now we look for the connection between the correlation functions and the
transport coefficients. 
First we consider the mobility.  With an applied external force
 $\overline{\bf K}(t) =\overline{\bf K}[\omega] e^{i\omega t} $ 
the mobility $\mu[\omega]$ is defined by 
$<{\bf u}_i [\omega]> = \mu_{ij}[\omega] \overline{\bf K}_{j}[\omega]$. 
From Eq.(1) we immediately obtain the mobility
\[
  \mu[\omega] = (im\omega + \eta[\omega] -i \sigma_y B )^{-1}
\]
in the limit $t_0\rightarrow -\infty$.
Using Eq.(4), the mobility is related to the velocity-velocity 
correlation function 
\be
  \mu[\omega] = \frac{{\cal U}[\omega] }{k_B T} \, .
\ee 
This is the `first' fluctuation-dissipation theorem described by 
Kubo\cite{kubo},
equivalent to the Nakano-Kubo's formula for
the electrical conductivity.

Next we consider that the particle is moving at a given 
velocity $\overline{\bf u}(t)$ and find out what is the 
external force needed to sustain such a motion. 
It is equivalent to the calculation of
resistivity if the particle is charged. 
From Eq.(1), we have the average force 
\be
   < K_i[\omega] > 
 = (im \omega + \eta[\omega] - i \sigma_y B )_{ij}
          \; \overline{u}_{j}[\omega]  
\ee
which is trivially equivalent to the reciprocal of
conductivity formula. Obviously this process does not provide us an
independent way of calculating resistivity.

However, if we are only interested in the average force  
$ <{\bf K}> $ in a steady state motion, 
we do have an alternative resistivity formula. 
After  taking $ \omega\rightarrow 0$ and using Eqs.(6) and (8),
Eq.(10) gives
\be 
   < K_i >[0]
    = \frac{ 1 }{k_B T}  \left( {\cal R}_{ij}[0]
         - {\cal F}_{ij}[0]  \right) \overline{u}_j[ 0 ]  \, .
\ee
Taking  $\eta$ to be a scalar(proportional to a unit matrix), 
the external force can take a more suggestive form,
\be
   <{\bf K}[0]> = \eta[0] \; \overline{\bf u}[0] 
                - \overline{\bf u}[0]\times {\bf B} \; ,
\ee
where the longitudinal component depends on ${\cal R}[0]$, 
the random force-force correlation function, 
and the transverse component only on ${\cal F}[0]$, the total force-force 
correlation function.  
Eq.(11) is the steady state resistivity formula. 
It provides a direct way to obtain
DC resistivity from force correlation functions.  
The straight forward interpretation of Eq.(12) is the force-balance: 
The externally applied force to keep the steady velocity
is equal in magnitude but opposite in sign 
to the sum of the frictional and the Lorentz forces.
Above analysis  shows that the transverse
force is not affected  by the thermal reservoir under the assumption that
$\eta[\omega]$ is a scalar.

So far, all of our calculations are exact. 
Now we will discuss how the results may 
change when employing the relaxation time
approximation to account for the
existence of thermal reservoir in comparison with  the exact calculations.  
Without the thermal reservoir, the velocity-velocity
correlation is given by
\be
   {\cal U}[\omega] = (i m \omega -i \sigma_y B )^{-1}  
     m \; {\cal U}(0) \; .
\ee
Then  we switch on the thermal reservoir to allow the relaxation
process to happen. We use a relaxation time
approximation by the standard rule,
$i\omega\rightarrow i\omega + \eta [\omega]/m$. 
Substituting it into
Eq.(13), we have found the velocity-velocity
correlation under the relaxation time approximation is given by
\[
  {\cal U}[\omega]
  = (i m \omega + \eta[\omega] -   i \sigma_y B )^{-1}  k_B T\; ,
\]
which is exactly the same as the one obtained by above rigorous
calculation.
We conclude that the relaxation time approximation can be a valid one
for velocity-velocity correlations when used in a conductivity formula.

Next we evaluate the force-force correlation by the relaxation time 
approximation.
Without  thermal reservoir, 
the random force correlation is zero, that is, ${\cal R}(t) = 0$.
If we switch on the thermal  reservoir by using a relaxation time 
approximation $  i\omega \rightarrow i\omega + \eta[\omega]/m$,
the random force correlation is still
incorrectly set to zero, and cannot be made finite. 
The total force correlation without thermal reservoir is
\be
   \ba{ll}
    {\cal F}_{ij} [\omega]  =  &
  \left( i B\sigma_y  + i m \omega + (m\omega)^2 \right. \\
    & \left. (i m \omega -i \sigma_y B )^{-1}
        \right)_{ik} m {\cal U}_{kj}(0)  \, . \\
  \ea
\ee
When we switch on the thermal  reservoir by using a relaxation time 
approximation $  i\omega \rightarrow i\omega + \eta[\omega]/m$ 
in Eq.(14), we have 
\[
  \ba{ll}
  {\cal F} [\omega]  =  & 
   \left( iB\sigma_y + i m \omega + \eta[\omega]
   - (im \omega + \eta[\omega])^2 \right. \\
  & \left. \left( i m \omega  +   \eta[\omega] 
  -i \sigma_y B \right) ^{-1}
  \right) k_B T  \,  .
  \ea
\]
This is a rather complicated expression. 
We can simplify it in the limit $\omega<< \eta[\omega]$,
or $ \omega \tau<<1 $, $\tau=m/\eta[0] $ is a relaxation time:
\be
   {\cal F}[0]
    = \frac{B }{ 1+ (\omega_0 \tau)^2 } \left(
      \omega_0 \tau +i\sigma_y (\omega_0 \tau)^2 \right) k_B T \; , 
\ee
with 
$\omega_0 = B/m$.
Let us use the resistivity formula Eq.(11) to calculate the external
force needed to keep the particle moving with a given 
velocity. With ${\cal R}[\omega] = 0$ and  ${\cal F}[0]$ given by Eq.(15),
the external force is
\be
   <{\bf K}[0]> =   
     - \frac{\omega_0 \tau}{ 1+(\omega_0 \tau)^2} ( B
     \overline{\bf u}[0]   +   
     \omega_0\tau \; \overline{\bf u}[0] \times {\bf B} )  \, . 
\ee
These results have no connection at all to the rigorous
results shown in Eq.(12). Evidently the relaxation time
approximation cannot be valid in such a calculation.

However, Eq.(16) is familiar to us. 
With merely a re-definition of constants $B=\kappa\rho$ with
$\kappa$ the circulation and $\rho$ the superfluid density, 
and $\omega_0$ as the the core level spacing,
this force becomes exactly the same as 
the one appeared in the derivation of vortex dynamics using the  
relaxation time approximation\cite{volovik,simanek,kp2,vo,stone},
where the average is made for all other degrees of freedoms except
th0se of the vortex.
The literal interpretation of such results is that
the Lorentz force  on a moving particle 
is canceled by the effect of thermal reservoir. 
It also shows a different friction\cite{fi}.
The similarity of structure between Eq.(16) and those obtained
in vortex dynamics with relaxation time approximation suggests
that they may have the same source of error.
The calculation  by \u{S}im\'{a}nek \cite{simanek}
has explicitly used the relaxation time approximation in
the force-force correlation function.
Although several other publications \cite{kp2,vo}
have not explicitly specified their methods as force-force correlation function
calculations, their final formula  are the same as
Ref. \cite{simanek}.

There are also exceptions.
In Ref.\cite{stone}, Stone made an attempt to use  Boltzmann equation 
to solve the forces on a moving vortex.
When we check Eq. (5.8) in Ref.\cite{stone}, we find that
if we substitute $<{\bf k}> = <{\bf k}>_0$ into this equation,
where $<{\bf k}>_0$ is the equilibrium value of $<{\bf k}>$
in the frame where vortex is stationary, the equation 
is not satisfied. However, $<{\bf k}>$ should relax to $<{\bf k}>_0$ 
in such a case because  there is
no extra applied pinning force and the vortex is stationary in
this frame.
There are also  additional problems.
The equation itself is not Galilean invariant.
Nonetheless, its solution has been transformed 
into lab frame using
Galilean invariance in order to obtain force on a moving vortex . 

Before we conclude, we return to fluctuation-dissipation theorems
and Berry's phase.
The type of fluctuating forces we use 
in stochastic process will not generate any addition transverse
force in low frequency limit because of their vanishing
off-diagonal correlation. However, even though they are
widely used, we may still question
the validation of these fluctuating forces. 
Our understanding is that when we separate a fluctuating
force from what we leave to be systemic, we 
generally choose to assign the average effect to  the later.
Thus the fluctuating force no longer has a zero frequency
component of off-diagonal correlation.
The  Berry's phase is  more powerful 
in this respect. If indeed we have left an off-diagonal correlation
with non-zero mean to 'random force', the zero frequency 
component will be included if we evaluate Berry's phase. 
Finally, we point out that when correctly evaluated, 
the core state transitions reproduces the results of 
Berry's phase calculation at zero temperature with impurities 
included, and the friction arises naturally\cite{az,az2}.

\noindent 
Acknowledgments {\ } {\ }
{ We thank Prof. Thouless for calling our attention to 
 Ref.\cite{bh} and Prof. Ballentine and Prof. Stone
 for informative discussions.
 We appreciate the hospitality of the Physics Department(XMZ) 
 and the Institute for
 Nuclear Theory(PA) at University of Washington.
 This work was financially supported in part by USA DOE(PA)
 and  Swedish NFR. }

\end{document}